\documentclass[10pt,a4paper,twoside]{article}

\usepackage{indentfirst}			                      
\usepackage{graphicx}			                 	        
\usepackage{amsgen,amsfonts,amssymb,amsmath,amsbsy}	

\newenvironment{resum}{\begin{quote}\small}{\end{quote}}

\newcommand{\bfsf}[1]{\textsf{\textbf{#1}}}

\newcommand{\be}{\begin{equation}}  
\newcommand{\ee}{\end{equation}}

\newcommand{\bea}{\begin{eqnarray}}           
\newcommand{\eea}{\end{eqnarray}} 
\newcommand{\ba}{\begin{align}}
\newcommand{\ea}{\end{align}}

\def\be{\begin{equation}}
\def\ee{\end{equation}}
\def\beq{\begin{eqnarray}}
\def\eeq{\end{eqnarray}}
\newcommand{\beqn}{\begin{eqnarray*}}
\newcommand{\eeqn}{\end{eqnarray*}}
\def\lm{{\ell m}}

\def\l{{\ell }}

\begin{document}

\thispagestyle{plain}		

\begin{center}


{\LARGE\bfsf{Fluid accretion onto relativistic stars\\
and gravitational radiation}}

\bigskip


\textbf{Alessandro Nagar}$^{1,2}$ and \textbf{Guillermo Diaz}$^2$


$^1$\textsl{Universit\`a di Parma, Italy.} \\
$^2$\textsl{Departament d'Astronomia i Astrof\'isica, Universitat de Val\`encia, Spain.}

\end{center}

\medskip


\begin{resum}
This article reports results from numerical simulations of the gravitational 
radiation emitted from nonrotating relativistic stars as a result of the 
axisymmetric accretion of layers of perfect fluid matter, shaped in the 
form of quadrupolar shells. We adopt a {\em hybrid} procedure where we 
evolve numerically the polar nonspherical perturbations equations of the 
star coupled to a fully nonlinear hydrodynamics code that calculates the 
motion of the accreting matter. Self-gravity of the accreting fluid as 
well as radiation reaction effects are neglected.  
\end{resum}

\bigskip


\section{Introduction}
In this article we shall report of the work made in collaboration
with J.A.~Pons and J.A.~Font (see Ref.~\cite{nagar} for 
further details). The purpose is to analyze the gravitational wave 
emission pattern from neutron stars as a result of  
accretion of matter which is shaped in the form of quadrupolar
shells. Accretion is expected to happen following the gravitational
collapse of the core of a massive star, once a neutron star has 
already been formed. Part of the remaining stellar material which has 
not been expelled by the shock driving the supernova explosion may fall back 
onto the neutron star, until a critical mass is exceeded and the star collapses 
to a black hole. Some more material may in turn form a long-lived, 
centrifugally-supported torus or disk if the collapsing star had initially 
some amount of rotation \cite{daigne}. A detailed realistic modelization 
of the gravitational emission from accretion flows would require  
three-dimensional (magneto-) hydrodynamical simulations in general 
relativity, coupled to radiation transport and diffusive processes. However, 
some preliminary steps can be taken to understand the underlying basic 
physics in a qualitative way. The procedure adopted here lies in the borderline of 
full numerical relativity and perturbation theory and extends to relativistic stars 
the 2D-axisymmetric computations concerned with black holes, formerly performed in Ref. \cite{PapadopoulosFont}.
The approach is as follows: the accreting matter, whose self-gravity is neglected, 
is evolved in a curved static background by solving the nonlinear hydrodynamics equations;
the response of the star to the infalling matter, which triggers the emission of gravitational 
radiation, is computed by numerically solving the metric perturbations equations, with 
sources accounting of the fluid motions. The key assumption of this method is that the mass of 
the accreting fluid is much smaller than that of the star, so that, fluid {\it self-gravity} 
as well as {\it radiation reaction} effects are neglected. The first approximation (no self-gravity) 
is in general valid for fluid motions in the vicinity of the star, where tidal forces 
dominate over the fluid self-gravity. The second approximation (no radiation reaction) is 
valid as long as the energy in the form of gravitational radiation is much smaller than 
the kinetic or internal energy of the fluid. We use then a particular implementation of 
the theory of metric perturbations of stars (see Ref.~\cite{KokkotasSchmidt} and references 
therein for an extended description). A further simplification adopted is that the star 
is non rotating, so that metric, pressure, energy density and mass are obtained as 
solution of the Tolman-Oppenheimer-Volkoff equations of stellar equilibrium \cite{MTW}. 
We assume barotropic equation of state written in polytropic form $p=K\epsilon^{\Gamma}$ 
to model a cold neutron star. We show results for two stellar models of mass $M=1.4M_{\odot}$ 
with $\Gamma=2$, the one being more compact than the other. The more compact model (A), 
has central density $\epsilon_c=2.455 \times 10^{15}$ g/cm$^3$, and $K=122.25$ km$^2$, 
with $R=9.80$ km. The less compact model (B) has $\epsilon_c=0.92\times 10^{15}$g/cm$^3$, 
and $K=180$ km$^2$ which leads to $R=13.44$ km.
\section{Even-parity perturbations of a relativistic star}
The presence of a fluid distribution outside the star creates a small perturbation 
in its gravitational field, so that the total metric is expressed as 
$g_{\mu\nu}=g_{\mu\nu}^0+\delta g_{\mu\nu}$, where $g_{\mu\nu}^0$ is the metric 
of the star and $\delta g_{\mu\nu}$ is the perturbation;
this can be decomposed in odd--parity (axial) and even--parity (polar) modes, each 
carrying spherical-harmonic indexes $\l$ and $m$~\cite{KokkotasSchmidt}. Since the
hydrodynamics evolution is 2D-axisymmetric ($\partial_{\varphi}=0$) and because the 
star and the shell have no angular momentum, only the polar modes are excited by 
the accreting matter. For this reason we shall restrict the discussion below just 
to polar perturbations of stars. Formulations of the polar perturbations equations 
of a nonrotating stars suitable for time evolution are currently available in literature 
(see Ref.~\cite{nagar} and references therein). Our calculations are performed in the 
Regge-Wheeler gauge \cite{ReggeWheeler}, by specializing the gauge-invariant and coordinate 
independent formalism of Ref.~\cite{Gundlach}. The equations obtained are equivalent 
to those of Ref.~\cite{allen} and \cite{RuoffI}, although different metric variables are used. 
Thus, for each $(\l,m)$ pair, $\delta g_{\mu\nu}$ is parametrized by two scalar 
quantities, $k$ (the perturbed 3-conformal factor) and  $\chi$ (the actual gravitational 
wave degree of freedom), and it reads
\begin{align}
\delta g_{\mu\nu}dx^{\mu}dx^{\nu}=\left[(\chi+k)e^{2a}dt^2-2\psi e^{a+b}dtdr+(\chi+k)e^{2b}dr^2+kr^2d\Omega^2\right]Y^{\lm}\;.
\end{align}
On a static background $\psi$ does not play any dynamical role, as it can be 
obtained by quadrature from $k$ and $\chi$ \cite{Gundlach}. We note that, 
despite this problem has actually two degrees of freedom inside the star and 
one outside it, we have found more convenient for numerical reasons to consider 
an additional variable inside the star, the perturbation of the relativistic 
enthalpy $H=\delta p/(p+\epsilon)$. We are left, then, with three degrees of 
freedom inside the star ($\chi$, $k$ and $H$) and two outside ($\chi$ and $k$). 
Outside the star, from $k$ and $\chi$ we get the Zerilli-Moncrief function 
\cite{Zerilli,Moncrief}
\begin{equation}\label{DefZeta}
Z = \dfrac{4 r^2 e^{-2b}}{\lambda\left[(\lambda-2)r+6M\right]}
\bigg[\chi+\left(\frac{\lambda}{2}+\frac{M}{r}\right)e^{2b}k-r~k_{,r}\bigg],
\end{equation}
with $\lambda=\l(\l+1)$, whose Fourier transform gives the energy radiated in GWs 
\begin{align}
E^{\lm}=\dfrac{\pi}{8}
\,\dfrac{(\l+2)!}{(\l-2)!}\int_0^{\infty}f^2\left|\tilde{Z}(f,r)\right|^2df\;.
\end{align}
%
%
\begin{figure}[t]
\begin{center}
\includegraphics[width=60 mm,height = 49 mm]{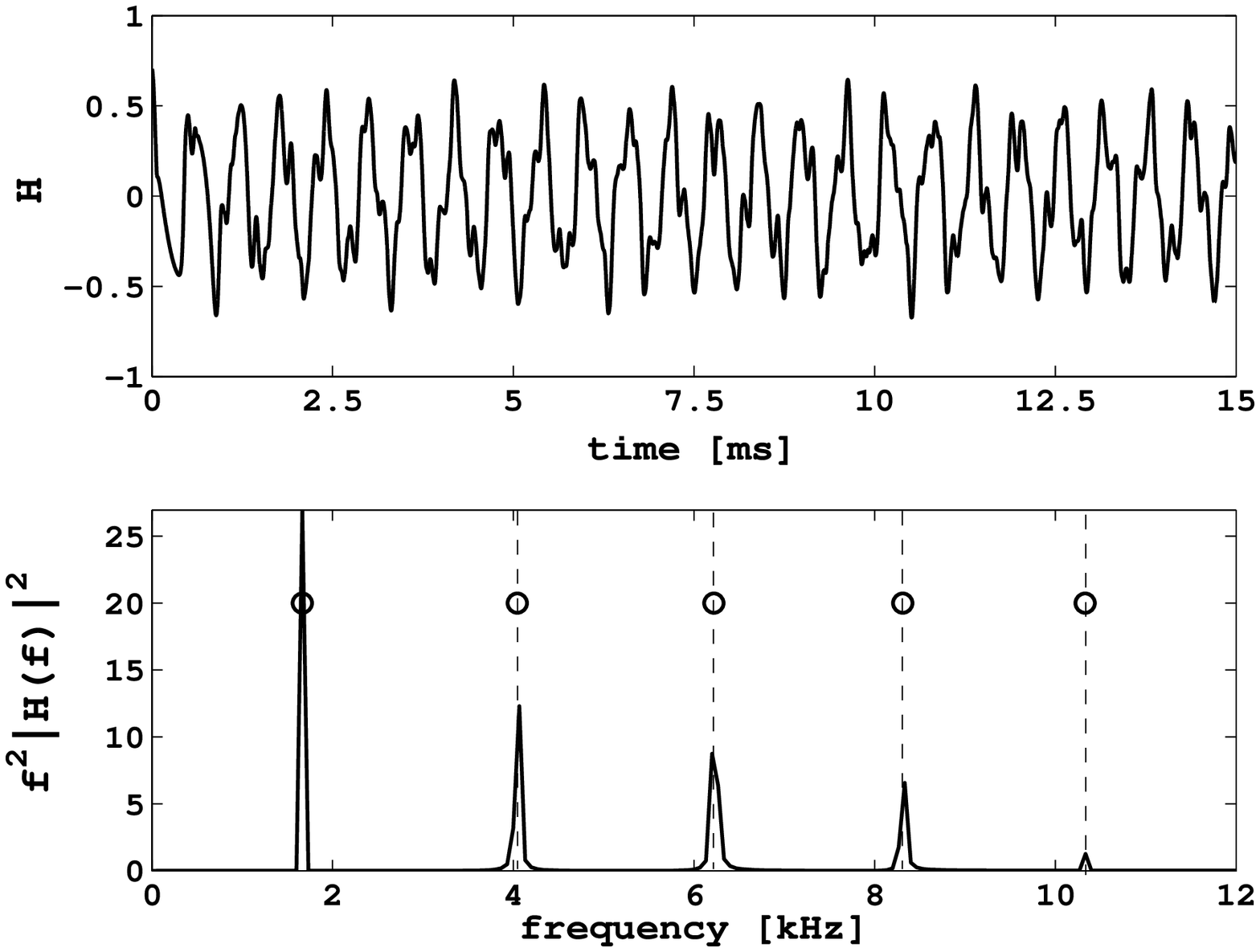}\qquad
\includegraphics[width=60 mm,height = 49 mm]{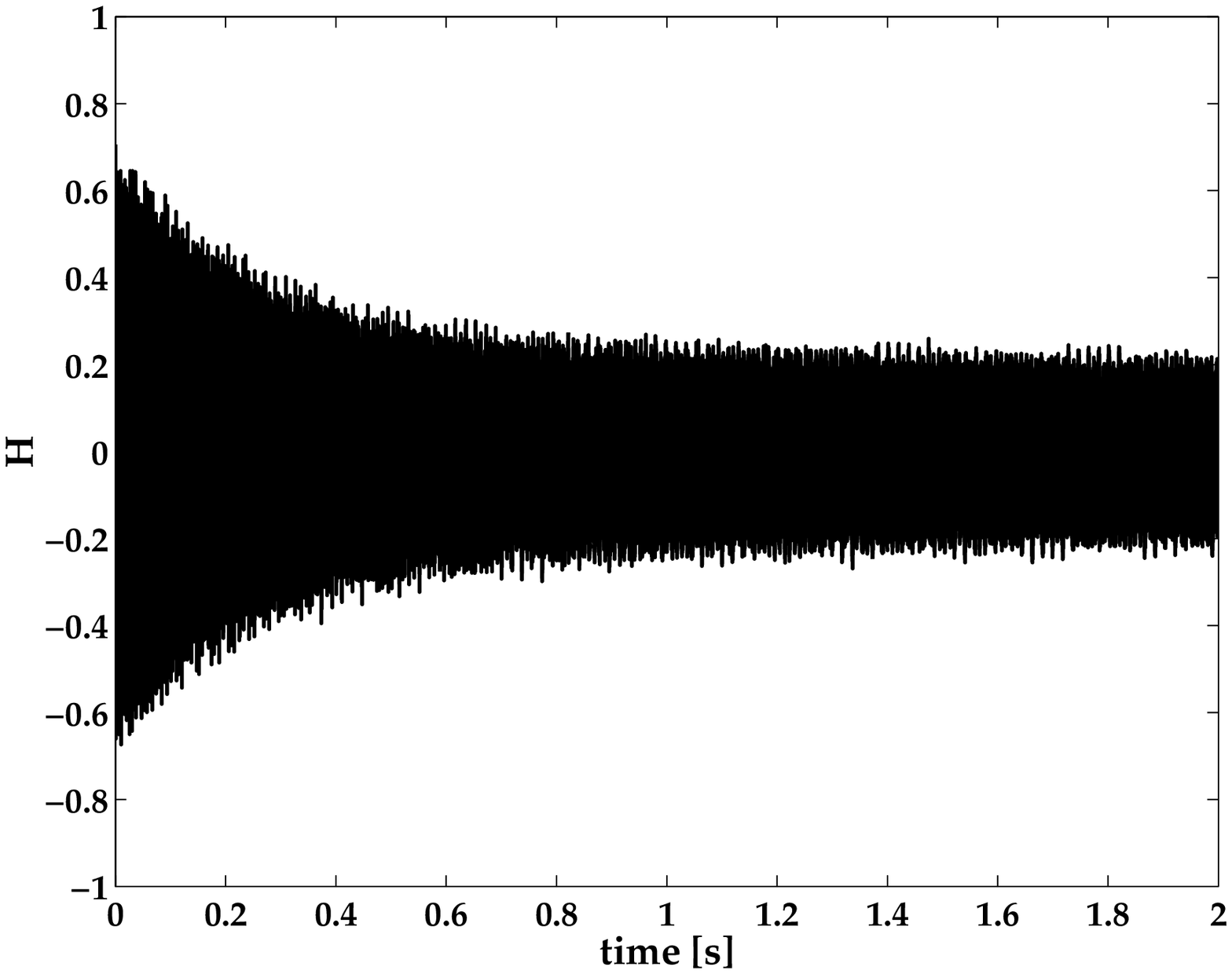}
\caption{\label{label:fig1}Time evolution of the enthalpy perturbation of model B}
\end{center}
\end{figure}
%
%
Unlike other studies \cite{allen,RuoffI}, we choose to evolve the perturbations 
through a couple of hyperbolic equations for $\chi$ and $H$ plus an elliptic 
equation, the Hamiltonian constraint, which is solved for $k$ at every temporal slice.
With this ``constrained'' scheme, that works as a dynamical corrector of the 
truncation errors, we gain two important benefits: on one hand we can obtain 
the frequencies of the fluid modes (by Fourier transforming the waveforms) with 
very good accuracy even with a relatively low resolution; on the other hand, we do 
not have limits to the evolution time (unlike free hyperbolic formulations, 
where sooner or later a violation of the constraints occurs and the evolution
must be stopped), and we can see dynamically the damping of the fluid modes due
to GWs emission. Fig.~\ref{label:fig1} shows the evolution of an initial enthalpy 
profile as $H=\sin(\pi r/2R)$ for model B, observed at half of the star radius. 
The circles in the Fourier spectrum represent the fluid 
modes obtained using a frequency-domain code based on the Lindblom-Detweiler 
formulation of the perturbed equations \cite{LindblomDetweiler} and described 
in \cite{PonsStructure}. The relative difference between the location of the peaks 
and these values is of some parts in a thousand, with a resolution of 200 points. 
The right panel of Fig.~\ref{label:fig1} refers to the same evolution carried out
for $2$ s. The damping time of the $f$-mode for this model is $\tau_f=0.26$ s:
after $\sim 1$ s just the longer living $p$-modes survive in the signal \cite{KokkotasSchmidt}. 
\section{Gravitational waves from fluid accretion}
We present an overview of a typical simulation of accretion onto model A. Initially, the non 
spherical density profile is surrounded by a background fluid (an ``atmosphere'') 
satisfying the stationary and spherically symmetric Michel solution~\cite{michel}. 
The initial rest mass density profile is given by
\begin{align}
\rho=\rho_{\mathrm{Mich}}+\rho_{\mathrm{max}}e^{-\kappa(r-r_0)^2}\sin^2\vartheta\;,
\end{align}
with $\rho_{\mathrm{Mich}}$ being the profile consistent with the Michel solution. 
The mass of the shell is $\mu=0.01M$, which corresponds to a maximum density of 
$\rho_{\mathrm{max}}\sim 3.5\times 10^{-6}$~$\mathrm{km}^{-2}$ when we fix its 
width to $\kappa=1$. The (inhomogeneous) density profile of the atmosphere $\rho_{\mathrm{Mich}}$ 
is roughly three orders of magnitude lower than $\rho_{\mathrm{max}}$. The shell obeys a 
polytropic ($p={\cal K}\rho^{\gamma}$) EoS with ${\cal K}=0.01$ $\mathrm{km}^{2/3}$ 
and $\gamma=4/3$. The internal energy profile is obtained from the baryonic density $\rho$ 
and $p$ through the first law of thermodynamics as $e=p/((\gamma-1)\rho)$. The shell is 
initially at rest at $r_0=20$ km, and the waveform is extracted at $r_{\mathrm{obs}}=250$ km.

The motion of the fluid in a curved spacetime is governed by the local conservation laws
of baryonic number and energy-momentum; as shown by \cite{Banyuls}, the equations of general
relativistic hydrodynamics can be written as a hyperbolic system of balance laws with sources,
and then solved using a Godunov-type scheme based on the characteristic information of the 
system. The hydrodynamics code employed here is the same of Ref.~\cite{PapadopoulosFont}. 

In a realistic scenario, such as e.g.~fall-back accretion after a supernova explosion, 
the accreting matter interacts with the neutron star until a critical mass is reached and 
the star collapses to a black hole. As we cannot model this complex phenomenon within the 
current perturbative framework, we impose reflecting boundary conditions at 
the inner edge of the hydrodynamics domain (i.e.~the surface of the star), so that it is 
seen by the external fluid as a {\it hard} sphere. This choice includes the most relevant 
effect, that is, the pressure gradient stops the infalling matter, and the accretion 
process is then followed by the formation of shock waves which propagate off the stellar 
surface. The impact of the shell perturbs the star and triggers its quasinormal modes 
of pulsation, so that it radiates gravitational waves.

\begin{figure}[t]
\begin{center}
\includegraphics[width=60 mm,height = 49 mm]{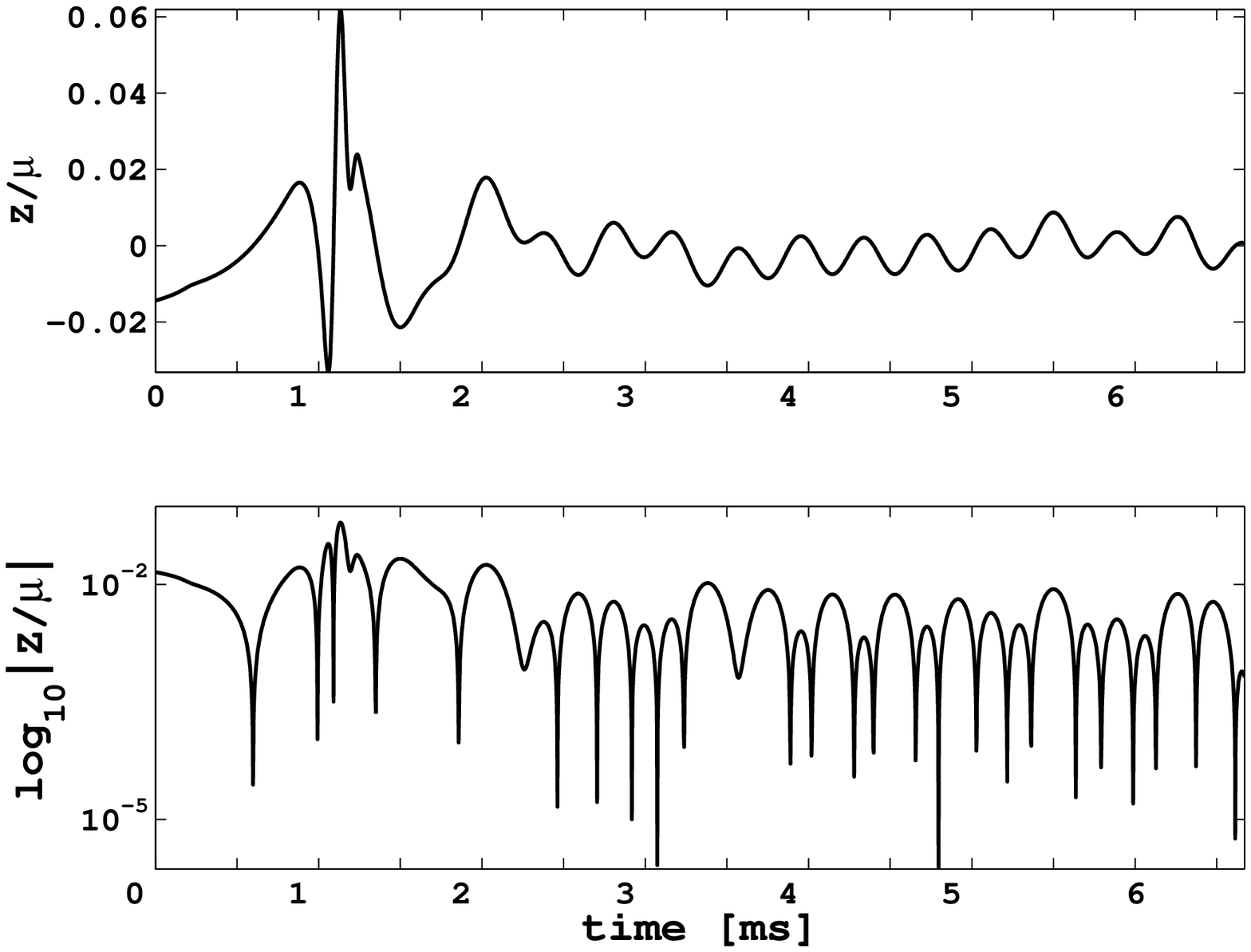}\qquad
\includegraphics[width=60 mm,height = 49 mm]{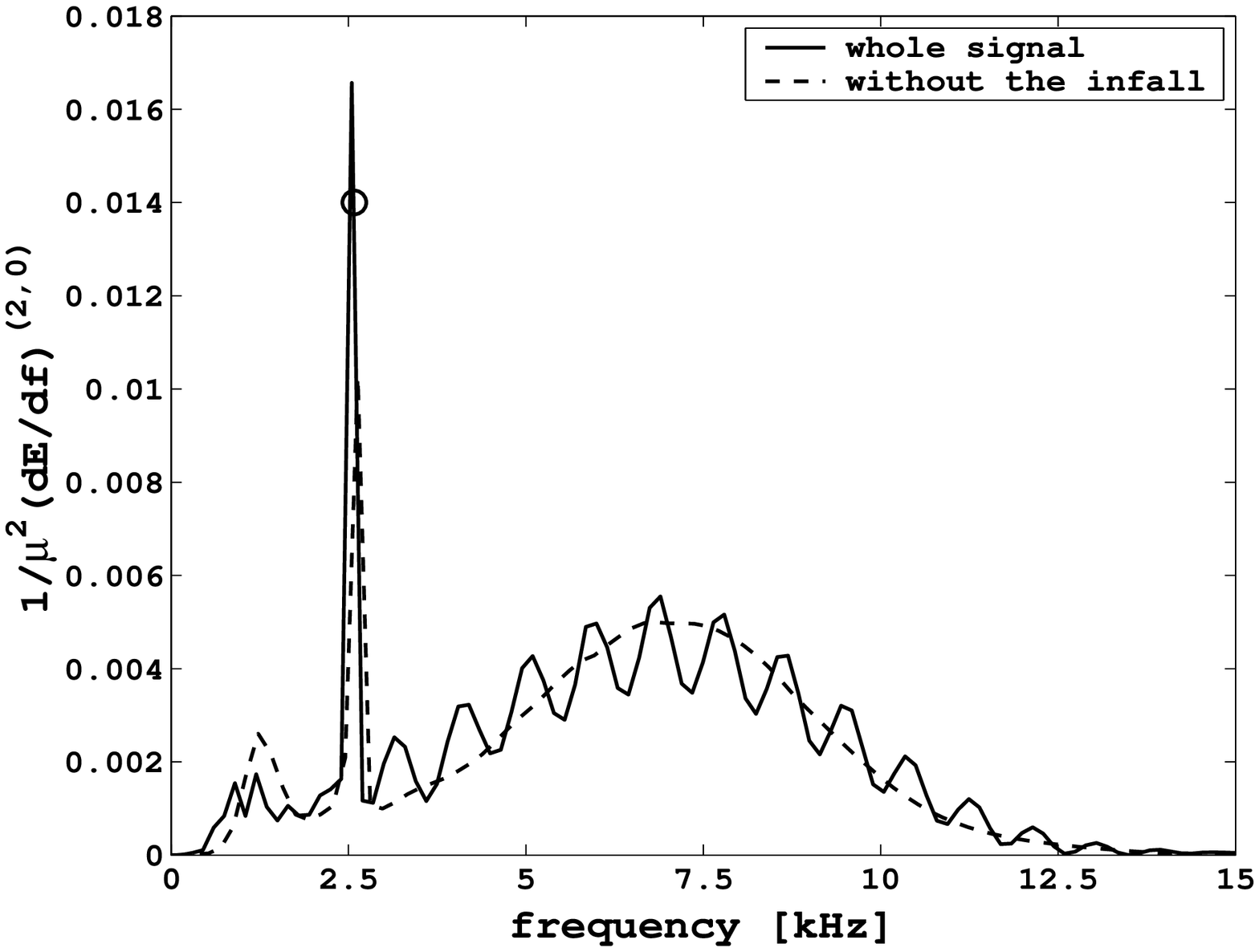}
\caption{\label{label:fig2}Waveform and energy spectra from a fluid shell falling 
on to star model A.}
\end{center}
\end{figure}

Fig.~\ref{label:fig2} exhibits the time evolution of the Zerilli-Moncrief function and the 
corresponding energy spectrum. In the waveforms, three phases can be identified. 
Firstly, the infalling, when the bulk of the shell is evolving outside the star, 
gradually approaching it, which is characterized by the steady increase of the amplitude 
of the signal: it is very short as the shell is initially located very close to 
the stellar surface. Secondly, a burst--like peak appears in the GW signal, which, 
as found in simulations of gravitational core collapse~\cite{Dimmelmeier}, coincides 
with the moment when the shell reaches the surface, creating a shock wave which 
propagates off the surface. Finally, the ringdown phase, characterized by a GW signal 
which is not exactly monochromatic as a result of the complex interaction between the 
gravitational field of the star and the layers of fluid captured on top of the stellar 
surface in the process of readjusting themselves to a new stationary solution. 
We note that, despite a remarkable resemblance with the results of core collapse 
simulations~\cite{Dimmelmeier}, there are also important differences in the post-bounce 
phase dynamics, as the ringdown of the neutron star lasts for much longer times. 
In our idealized setup these pulsations are not quickly damped by the existence of a 
dense envelope surrounding the star, as happens in the core collapse situation.
%
%
%
%
Concerning the energy spectra, the solid line is obtained by Fourier transforming 
the complete signal; correspondingly, the dashed line comes from the truncated waveform, 
in which we only take into account the contribution from the beginning of the burst 
($\sim 0.9$ ms) onward. The whole spectrum shows a complex structure: first of all, 
we find a narrow peak, identified with the $f$-mode of the star with an error around 
1\%. At higher frequencies, we have a broad band spectrum modulated by roughly equally
spaced bumps ($\delta f\approx 0.9$ kHz). These are interference effects coming from
the pre-bounce phase of the emission. This is confirmed by the fact that they are 
absent in the truncated spectrum. This latter presents one broad peak with a maximum at 
about $7$ kHz, which is however too low to be identified with the $w$-mode. As mentioned before, 
its origin should be related to the motion of the fluid shell and its interference with 
the gravitational field of the star: the high-frequency emission depends, then, on details 
of the dynamics of accretion rather than on intrinsic characteristics of the star. 
The total energy emitted in this process is computed by integrating the 
spectrum of Fig.~\ref{label:fig2}. We get $E^{20}\simeq3.02\times 10^{-8}\;M_{\odot}c^2\;$ or, 
in terms of the shell mass, $E^{20}\simeq 2.15\times 10^{-6}\;\mu c^2$.

As a last remark, we note that a small amplitude peak, at frequency lower than that of the 
$f$--mode, is present in the spectrum. It is associated with oscillations of that part of the 
external fluid that has been gravitationally captured by the star as a result of accretion. 
We argue that the existence of this unphysical low frequency peak is an artifact produced by 
the reflecting boundary conditions imposed at the surface. In a realistic scenario, the 
accreted matter would not simply bounce at the stellar surface, but it would rather interact 
with the neutron star envelope, resulting in heating and suffering nuclear reactions, until 
it is reabsorbed by the star.


\end{document}